\documentstyle[12pt,aasms4,psfig]{article}

\begin{document}
\title{\bf ZKDR Distance, Angular Size and Phantom Cosmology}

\author{R. C. Santos\altaffilmark{1} and J. A. S. Lima\altaffilmark{2}}
\affil{Departamento de Astronomia, Universidade de S\~ao Paulo,
\\
05508-900 S\~ao Paulo, SP, Brazil}
\altaffiltext{1}{cliviars@astro.iag.usp.br}
\altaffiltext{2}{limajas@astro.iag.usp.br}

\begin{abstract}
The influence of mass inhomogeneities on the angular size-redshift
test is investigated for a large class of flat cosmological models
driven by dark energy plus a cold dark matter component. The results
are presented in two steps. First, the mass inhomogeneities are
modeled by a generalized Zeldovich-Kantowski-Dyer-Roeder (ZKDR)
distance which is characterized by a smoothness parameter
$\alpha(z)$ and a power index $\gamma$, and, second,  we provide a
statistical analysis to angular size data for a large sample of
milliarcsecond compact radio sources. By marginalizing over the
characteristic angular size $l$, fixing $\Omega_M = 0.26$ and
assuming a Gaussian prior on $\omega$, i.e., $\omega = -1 \pm 0.3$,
the best fit values are $\omega = -1.03$ and $\alpha = 0.9$.  By
assuming a Gaussian prior on the matter density parameter, i.e.,
$\Omega_M = 0.3 \pm 0.1$, the best fit model for a phantom cosmology
with $\omega = -1.2$ occurs at $\Omega_M = 0.29$ and $\alpha = 0.9$
when we marginalize over the characteristic size of the compact
radio sources. The results discussed here suggest that the ZKDR
distance can give important corrections to the so-called background
tests of dark energy.
\end{abstract}
\pagebreak

\section{Introduction}

\hspace{0.5cm}An impressive convergence of recent astronomical
observations are suggesting that our world behaves like a spatially
flat scenario dominated by cold dark matter (CDM) plus an exotic
component endowed with large negative pressure, usually named dark
energy (Perlmutter et al. 1998;  Efstathiou {\it et al.} 2002; Riess
{\it et al.} 1999, 2004; Allen {\it et al.} 2004; Astier {\it et
al.} 2006; Riess {\it et al.} 2007). In the framework of general
relativity, besides the cosmological constant, there are several
candidates for dark energy, among them: a vacuum decaying energy
density, or a time varying $\Lambda(t)$ (Ozer \& Taha 1986, 1987;
Bertolami 1986; Freese {\it et al.} 1987; Carvalho {\it et al.}
1992; Lima \& Maia 1994; Lima \& Trodden 1996; Lima 1996; Torres \&
Waga 1996; Overduin \& Cooperstock 1998; Cunha \& Santos 2004;
Shapiro {\it et al.} 2005, Costa {\it et al.} 2007), the so-called
``X-matter" (Turner \& White 1997; Chiba {\it et al.} 1997, Alcaniz
\& Lima 1999, 2001, Cunha {\it et al.} 2003, D\c{a}browski 2007), a
relic scalar field (Peebles \& Ratra 1998; Caldwell {\it et al.}
1998; Ulam {\it et al.} 2004), and a Chaplygin Gas (Kameshchik {\it
et al.} 2001; Bil\'{\i}c {\it et al.} 2002; Bento {\it et al.} 2002;
Alcaniz \& Lima 2005). Some recent review articles discussing the
history, interpretations, as well as, the major difficulties of such
candidates have also been published in the last few years
(Padmanabhan 2003; Peebles \& Ratra 2003; Lima 2004; Turner \&
Huterer 2007).

In the case of X-matter, for instance, the dark energy component is
simply described by an equation of state $p_x = \omega\rho_x$. The
case $\omega = -1$ reduces to the cosmological constant,  and
together the CDM defines the scenario usually referred to as
``cosmic concordance model" ($\Lambda$CDM). The imposition $\omega
\geq -1$ is physically motivated by the classical fluid description
(Hawking \& Ellis 1973). However, as discussed by several authors,
such an imposition introduces a strong bias in the parameter
determination from observational data. In order to take into account
this difficulty, superquintessence or phantom dark energy
cosmologies have been recently considered where such a condition is
relaxed (Faraoni 2002; Caldwell et al. 2003; Gonzales-Diaz 2003,
Santos \& Alcaniz 2005, Linder 2007). In contrast to the usual
quintessence model, a decoupled phantom component presents an
anomalous evolutionary behavior. For instance, the existence of
future curvature singularities, a growth of the energy density with
the expansion, or even the possibility of a rip-off of the structure
of matter at all scales are theoretically expected (see, however, Alcaniz \& Lima 2004; Gonzalez-Diaz \& Siguenza (2004), de Freitas Pacheco \& Hovarth (2007) for a thermodynamic discussion). Although possessing such strange features, the phantom behavior is
theoretically allowed by some kinetically scalar field driven
cosmology (Chiba {\it et al}. 2000), as well as, by brane world
models (Shani \& Shtanov 2002, 2003, Wu {\it et al.} 2007), and,
perhaps, more important to the present work, a PhantomCDM cosmology
provides a better fit to type Ia Supernovae observations than does
the $\Lambda$CDM model(Alam et al. 2003; Choudury \& Padmanabhan
2004; Astier {et al.} 2006). Many others observational and
theoretical properties of phantom driven cosmologies have also been
successfully confronted to standard results (see, for instance,
Alcaniz 2004; Piao \& Zhang 2004; Choudury and Padmanabhan 2005;
Perivolaropoulos 2005; Nesseris \& Perivolaropoulos 2007).

In this context, one of the most important tasks for cosmologists
nowadays is to confront different cosmological scenarios driven by
cold dark matter (CDM) plus a given dark energy candidate with the
available observational data. As widely known, a key quantity for
some cosmological tests is the angular distance-redshift relation,
$D_{A}(z)$, which for a homogeneous and isotropic background, can
readily be derived by using the Einstein field equations for the
Friedmann-Robertson-Walker (FRW) geometry. 
From $D_{A}(z)$ one obtains the expression for the angular
diameter $\theta (z)$ (see section 3) which can be compared with the
available data for different samples of astronomical objects
(Gurvits {\it et al.} 1999; Lima \& Alcaniz 2000, 2002; Gurvits
2004; Alcaniz \& Lima 2005).

Nevertheless, the real Universe is not perfectly homogeneous, with
light beams experiencing mass inhomogeneities along their way.
Actually, from small to intermediate scales ($\leq 100$Mpc), there
is a lot of structure in form of voids, clumps and clusters which
is probed by the propagating light. Since the perturbed metric is
unknown, an interesting possibility to account for such an effect
is to introduce the smoothness parameter $\alpha$ which is a
phenomenological representation of the magnification effects
experienced by the light beam. From general grounds, one expects a
redshift dependence of $\alpha$ since the degree of smoothness for
the pressureless matter is supposed to be a time varying quantity
(Linder 1988). When $\alpha = 1$ (filled beam), the homogeneous
FRW case is fully recovered; $\alpha < 1$ stands for a defocusing
effect while $\alpha = 0$ represents a totally clumped universe
(empty beam). The distance relation that takes these mass
inhomogeneities into account was discussed by Zeldovich (1964)
followed by Kantowski (1969), although a  clear-cut application
for cosmology was given only in 1972 by Dyer \& Roeder (many
references may be found in the textbook by Schneider, Ehlers \&
Falco 1992; Kantowski 2003). Many studies involving the ZKDR
distances in dark energy models have been published in the
literature. Analytical expressions for a general background in the
empty beam approximation ($\alpha = 0$) were derived by Sereno
{\it et al.} (2001). By assuming that both dominant components may
be clustered they also discussed how the critical redhift, i.e.,
the value of $z$ for which $D_{A}(z)$ is a maximum (or $\Theta(z)$
minimum), and compared to the homogeneous background results as
given by Lima \& Alcaniz (2000), and, further discussed by Lewis
\& Ibata (2002), and Ara\'ujo \& Stoeger (2007). More recently, Demianski {\it et al.} (2003), derived an useful analytical approximate solution for a clumped
concordance model ($\Lambda$CDM) valid on the interval $0 \leq z
\leq 10$. Additional studies on this subject involving time delay
(Giovi \& Amendola 2001; Lewis \& Ibata 2002) gravitational
lensing (Kochanek  2002; Kochanek \& Schechter 2003) or even
accelerated models driven by particle creation (Campos \& Souza
2004) have also been considered.

Although carefully investigated in many of their theoretical and
observational aspects, an overview in the literature shows that a
quantitative analysis on the influence of dark energy in
connection with inhomogeneities present in the observed universe
still remains to be studied. Recently, the ZKDR distance was
applied for the $\theta(z)$ statistics with basis on a
$\Lambda$CDM cosmology with constant $\alpha$ (Alcaniz {\it et
al.} 2004). It was concluded that the best fit model occurs at
$\Omega_M = 0.2$ and $\alpha = 0.8$ whether the characteristic
angular size $\l$ of the compact radio sources is marginalized.

In this paper, we focus our attention on X-matter cosmologies with
special emphasis to phantom models ($\omega < -1$) by taking into
account the presence of a clustered cold dark matter. The mass
inhomogeneities will be described by the ZKDR distance
characterized by a smoothness parameter $\alpha (z)$ which depends
on a positive power index $\gamma$. The main objective is to
provide a statistical analysis to angular size data from a large
sample of milliarcsecond compact radio sources distributed over a
wide range of redshifts ($0.011 \leq z \leq 4.72$) whose distance
is defined by the ZKDR equation. As an extra bonus, it will be
shown that a pure CDM model ($\Omega_M = 1$) is not compatible
with these data even for the empty beam approximation ($\alpha =
0$). The manuscript is organized as follows. In section 2 we
derive the ZKDR equation. We also provide some arguments for a
locally nonhomogeneous Universe where the homogeneous contribution
of the dark matter obeys the relation $ \rho_h = \alpha
\rho_{o}(\rho_m /\rho_{o})^{\gamma}$ where $\gamma$ is a positive
number. In section 3 we analyze the constraints on the free
parameters $\alpha$, $\gamma$ and $\Omega_M$. We end the paper by
summarizing the main results in section 4.

\section{The Extended ZKDR Equation}

Let us now consider a flat FRW geometry ($c=1$)
\begin{equation}
ds^2 \ = \ dt^2 \ - \ R^2(t)\left(dr^2 + r^2d\theta^{2} +
r^{2}\sin^{2}\theta\,d\phi^{2}\right),
\end{equation}
where $R(t)$ is the scale factor. Such a spacetime is supported by
the pressureless CDM fluid plus a X-matter component of densities
$\rho_M$ and $\rho_x$, respectively. Hence, the total energy
momentum tensor, $T^{\mu\nu} = {T^{\mu\nu}}_{({M})} +
{T^{\mu\nu}}_{({x})}$, can be written as
\begin{equation}\label{EMT}
T^{\mu\nu} = [\rho_{M} + (1 + \omega)\rho_{x}] U^{\mu}U^{\nu} -
\omega\rho_{x} g^{\mu\nu},
\end{equation}
where $U^{\mu}=\delta^{\mu}_o$ is the hydrodynamics 4-velocity of
the comoving volume elements. In this framework, the Einstein
Field Equations (EFE)
\begin{equation}\label{EFE}
G^{\mu\nu}\equiv R^{\mu\nu} - \frac{1}{2}g^{\mu\nu}R = 8\pi
GT^{\mu\nu},
\end{equation}
take the following form:
\begin{equation}\label{FRW1}
({\dot{R} \over R})^{2} = H_{o}^{2}\left[\Omega_{\rm{M}}({R_{o}
\over R})^{3} + \Omega_x({R_{o} \over R})^{3(1 + \omega)}\right] ,
\end{equation}
\begin{equation}\label{FRW2}
{\ddot{R} \over R} = -{1 \over
2}H_{o}^{2}\left[\Omega_{\rm{M}}({R_{o} \over R})^{3} + (3\omega +
1)\Omega_x({R_{o} \over R})^{3(1 + \omega)}\right] ,
\end{equation}
where an overdot denotes derivative with respect to time and
$H_{o} = 100h {\rm{Kms^{-1}Mpc^{-1}}}$  is the Hubble parameter.
By the flat condition, $\Omega_x = 1-\Omega_{\rm{M}}$, is the
present day dark energy density parameter. As one may check from
(2)-(5), the case $\omega = - 1$ describes effectively the favored
``cosmic concordance model" ($\Lambda$CDM).

On the other hand, in the framework of a  comformally flat FRW
metric, the optical scalar equation in the geometric optics
approximation reads (Optical shear neglected)

\begin{eqnarray}\label{sachs}
{\sqrt{A}}'' +\frac{1}{2}R_{\mu \nu}k^{\mu}k^{\nu} \sqrt{A}=0,
\end{eqnarray}
where $A$ is the beam cross sectional area, plicas means
derivative with respect to the affine parameter describing the
null geodesics, and $k^{\mu}$ is a 4-vector tangent to the photon
trajectory whose divergence determines the optical scalar
expansion (Linder 1988; Giovi \& Amendola 2001; Demianski {\it et
al.} 2000, Sereno et al. 2001). The circular frequency of the
light ray as seen by  the  observer with 4-velocity $U^{\alpha}$
is $\omega= U^{\alpha}k_{\alpha}$, while the angular diameter
distance, $D_A$, is proportional to $\sqrt A$ (Schneider, Ehlers
\& Falco 1992).

As widely known, there is no an acceptable averaging procedure for
smoothing out local inhomogeneities. After Dyer \& Roeder (1972),
it is usual to introduce a phenomenological parameter, $\alpha
(z)=1-{\rho_{cl}\over <\rho_m>}$, called the ``smoothness"
parameter. For each value of $z$, such a parameter quantifies  the
portion of matter in clumps ($\rho_{cl}$) relative to the amount
of background matter which is uniformly distributed ($\rho_m$). As
a matter of fact, such authors examined only the case for constant
$\alpha$, however, the basic consequence of the structure
formation process is that it must be a function of the redshift.
Combining equations (2), (3) and (6), after a straightforward but
lengthy algebra one finds that the angular diameter distance,
$D_{A}(z)$, obeys the following differential equation
\begin{equation}\label{angdiamalpha}
 \left( 1+z\right) ^{2}{\cal{F}}
\frac{d^{2}D_A}{dz^{2}} + \left( 1+z\right) {\cal{G}}
\frac{dD_A}{dz} + {\cal{H}} D_A=0,
\end{equation}
which satisfies the boundary conditions:
\begin{equation}
\left\{
\begin{array}{c}
D_A\left( 0\right) =0, \\
\\
\frac{dD_A}{dz}|_{0}=1.
\end{array}
\right.
\end{equation}
The functions ${\cal{F}}$, ${\cal{G}}$ and ${\cal{H}}$ in
equation (7) read
\begin{eqnarray}
{\cal{F}}& =& \Omega_M (1+z)^3 + (1-\Omega_M)(1+z)^{3(\omega +1
)}\nonumber \\ \nonumber \\ {\cal{G}} &=& \frac{7}{2}\Omega_M
(1+z)^3 +\frac{3\omega +7}{2} (1-\Omega_M )(1+z)^{3(\omega
+1)}\nonumber \\ \nonumber
\\ {\cal{H}} &=& \frac{3\alpha(z)}{2}\Omega_M (1+z)^{3} +
\nonumber\\  && +\frac{3(\omega+1)}{2} (1-\Omega_M)(1+z)^{3(\omega
+ 1)}.
\end{eqnarray}
The smoothness parameter $\alpha(z)$, appearing in the expression
of ${\cal{H}}$, assumes the form below (see Appendix A for a
detailed discussion)
\begin{equation}
\alpha (z) = \frac{\beta_o(1+z)^{3\gamma}}{1 +
\beta_o(1+z)^{3\gamma}},
\end{equation}
where $\beta_o$ and $\gamma$ are constants. Note that the fraction
$\alpha_o = \beta_o/(1 + \beta_o)$ is the present day value of
$\alpha (z)$. In Fig. 1 we show the general behavior of
$\alpha(z)$ for some selected values of $\beta_o$ and $\gamma$.

\begin{figure}[t]
\vspace{.2in}
\centerline{\psfig{figure=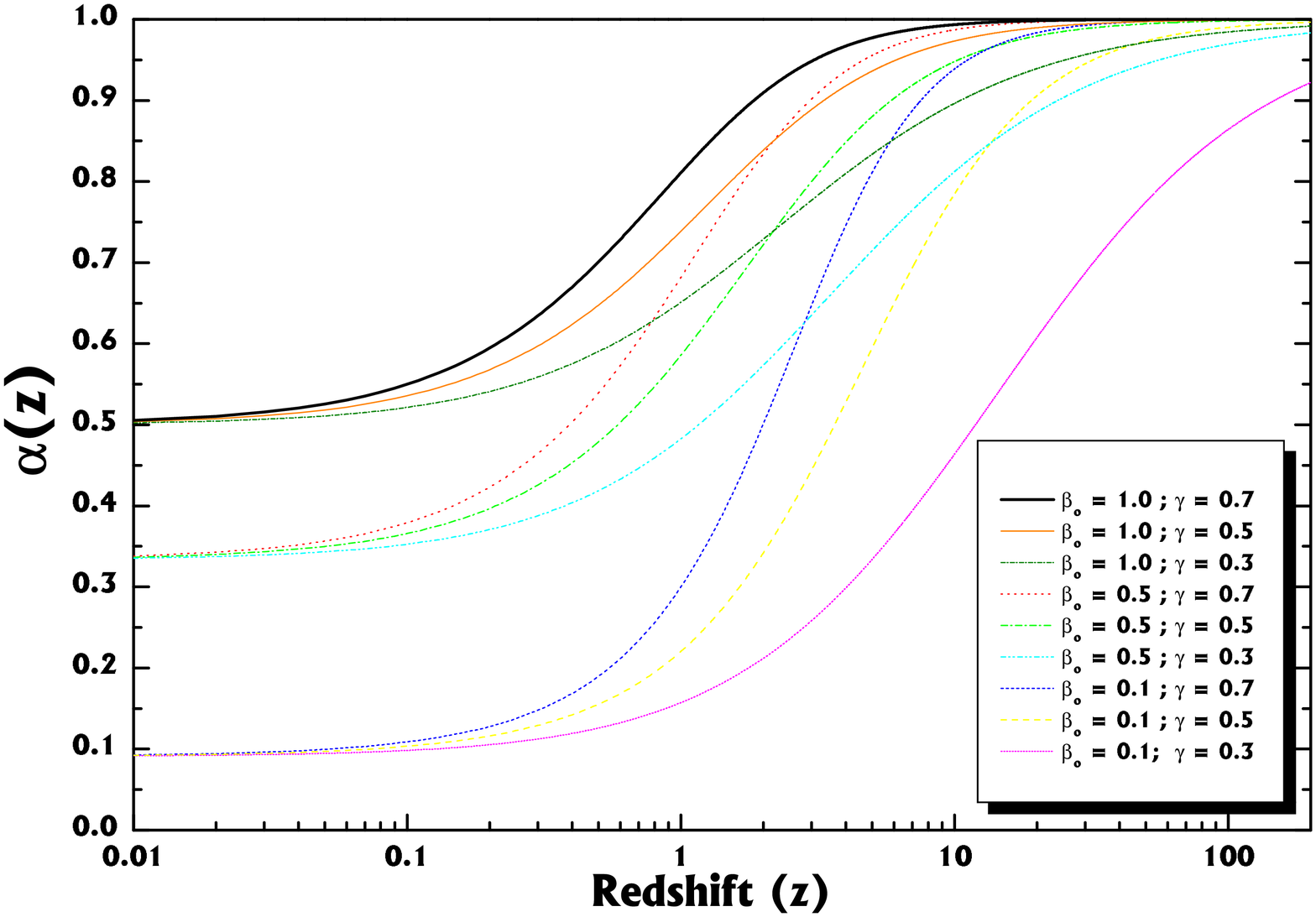,width=4.0truein,height=3.2truein}
\hskip 0.1in} \caption{The smoothness parameter as a function of the
redshift for some selected values of $\beta_o$ and $\gamma$. All
curves approach the filled beam result ($\alpha = 1$) at high
redshifts regardless of the values of $\beta_o$ and $\gamma$. Note
that $\beta_o$ determines $\alpha_o = \alpha(z=0)$. For a given
$\beta_o$ the curves starts at the same point but the rate
approaching unit (filled beam) depends on the $\gamma$ parameter.}
\end{figure}

At this point, it is interesting to compare Eq. (7) together the
subsidiary definitions (8)-(10) with  other treatments appearing
in the literature. For $\gamma = 0$ (constant $\alpha$) and
$\omega = -1$ ($\Lambda$CDM) it reduces to Eq. (2) as given by
Alcaniz {\it et al.} (2004).  In fact, for $\omega = -1$ the
function ${\cal{H}}$ is given by ${\cal{H}} =
\frac{3\alpha}{2}\Omega_M (1+z)^{3}$. A more general expression
for $\Lambda$CDM model (by including the curvature term) has been
derived by Demianski et al. (2004). As one may check, by
identifying $\omega \equiv m/3 -1$, our Eq. (7) is exactly Eq.(10)
presented by Giovi \& Amendola (2001) in their time delay studies
(see also Eq. (2) of Sereno et al. (2002)). It is worth notice
that in this paper the $\alpha$ parameter is greater than unity.
This means that the light rays are demagnified along the path (for
$\alpha < 1$ the light rays are magnified). In addition, the
$\alpha$ parameter may also depend on the direction along the line
of sight (for a discussion of such effects see Wang 1999, Sereno
et al. 2002).

Let us now discuss the integration of the ZKDR equation with
emphasis in the so-called phantom dark energy model ($\omega <
-1$). In what follows, assuming that $\omega$ is a constant, we
have applied for all graphics a simple Runge-Kutta scheme (see,
for instance, the rksuite package from www.netlib.org).

In Figure 2 one can see how the equation of state parameter,
$\omega$, affects the angular diameter distance. For  fixed values
of $\Omega_M = 0.3$, $\beta_o = 0.5$ and $\gamma = 0$, all the
distances increase with the redshift when $\omega$ diminishes and
enters in the phantom regime ($\omega<-1$). For comparison we have
also plotted the case for $\Lambda$CDM cosmology ($\omega=-1$).

\begin{figure}[t]
\vspace{.2in}
\centerline{\psfig{figure=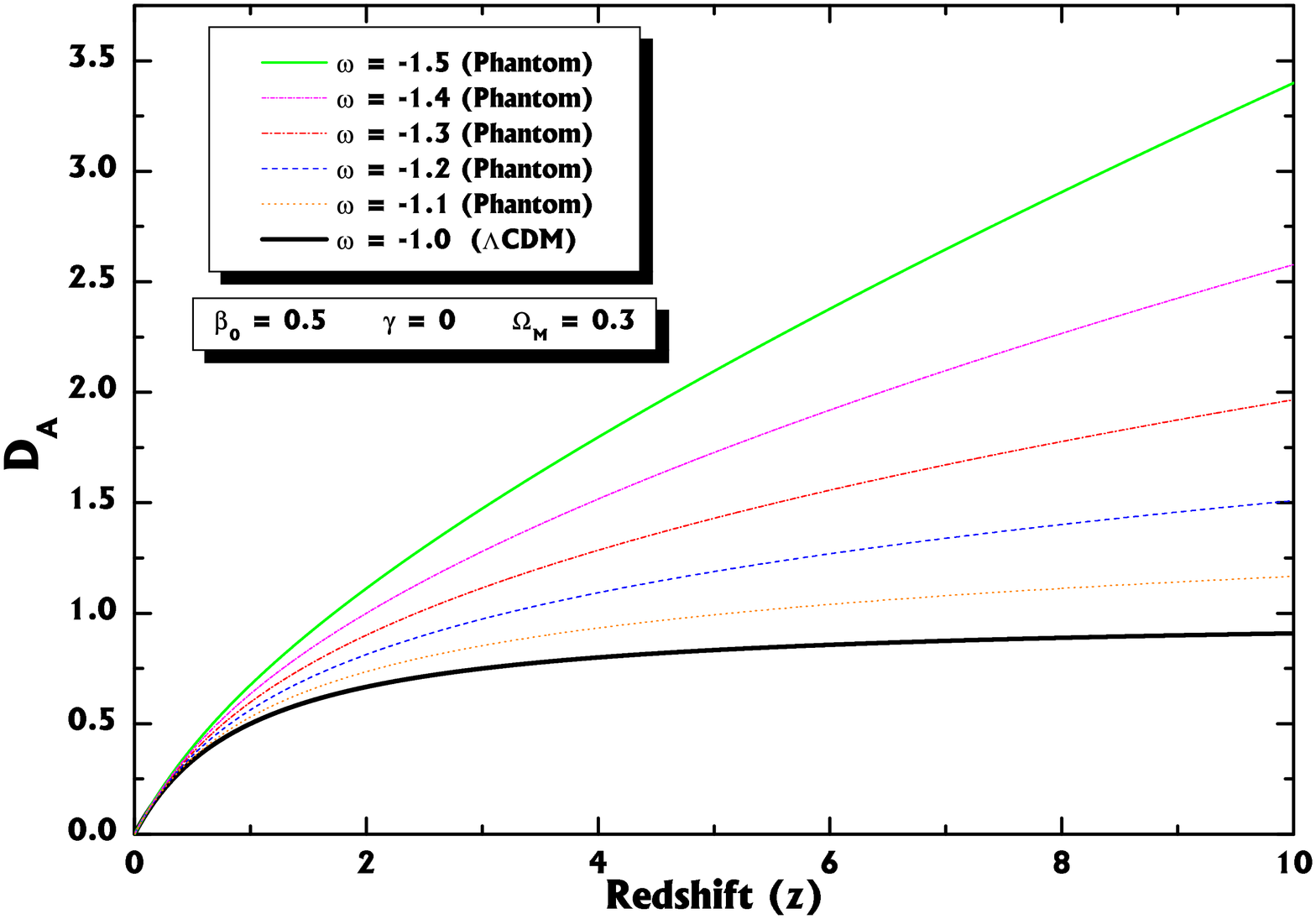,width=4.0truein,height=3.2truein}
\hskip 0.1in}\caption{Angular diameter distance for a flat FRW
phantom cosmology. The curves display the effect of the equation of
state parameter for $\beta_o=0.5$ and $\gamma = 0$. The thick curve
corresponds to the $\Lambda$CDM model. Note that for a given
redshift, the distances always increase for $\omega$ beyond the
phantom divide line ($\omega < -1$).}
\end{figure}

In  Fig. 3 we show the effect of the $\gamma$ parameter on the
angular diameter distance for a specific phantom cosmology with
$\omega = -1.3$, as requested by some recent analyzes of
Supernovae data (Riess 2004, Perivolopoulus 2004). For this plot
we have considered $\beta_o=0.5$. As shown in Appendix A, $\beta_o
= (\rho_h/\rho_{cl})_{z=0}$, is the present ratio between the
homogeneous ($\rho_h$) and the clumped ($\rho_{cl}$) fractions. It
was fixed in such a way that $\alpha_o$ assumes the value $0.33$.
Until redshifts of the order of 2, the distance grows for smaller
values of $\gamma$, and after that, it decreases following nearly
the same behavior.

In Fig. 4 we display the influence of the $\beta_o$ parameter on
the angular diameter distance for two distinct sets of $\gamma$
values. The cosmological framework is defined $\Omega_{M} = 0.3$
and the same equation of state parameter $\omega = -1.3$ (phantom
cosmology). For each branch (a subset of 3 curves with fixed
$\gamma$) the distance increases for smaller values of $\beta_o$,
as should be expected.

\section{ZKDR distance and Angular Size Statistics}

As we have seen, in order to apply the angular diameter distance
to a more realistic description of the universe it is necessary to
take into account local inhomogeneities in  the distribution of
matter. Similarly, such a statement remains true for any
cosmological test involving angular diameter distances,  as for
instance, measurements of angular size, $\theta(z)$, of distant
objects. Thus, instead of the standard FRW homogeneous diameter
distance one must consider the solutions of the ZKDR equation.

\begin{figure}[t]
\vspace{.2in}
\centerline{\psfig{figure=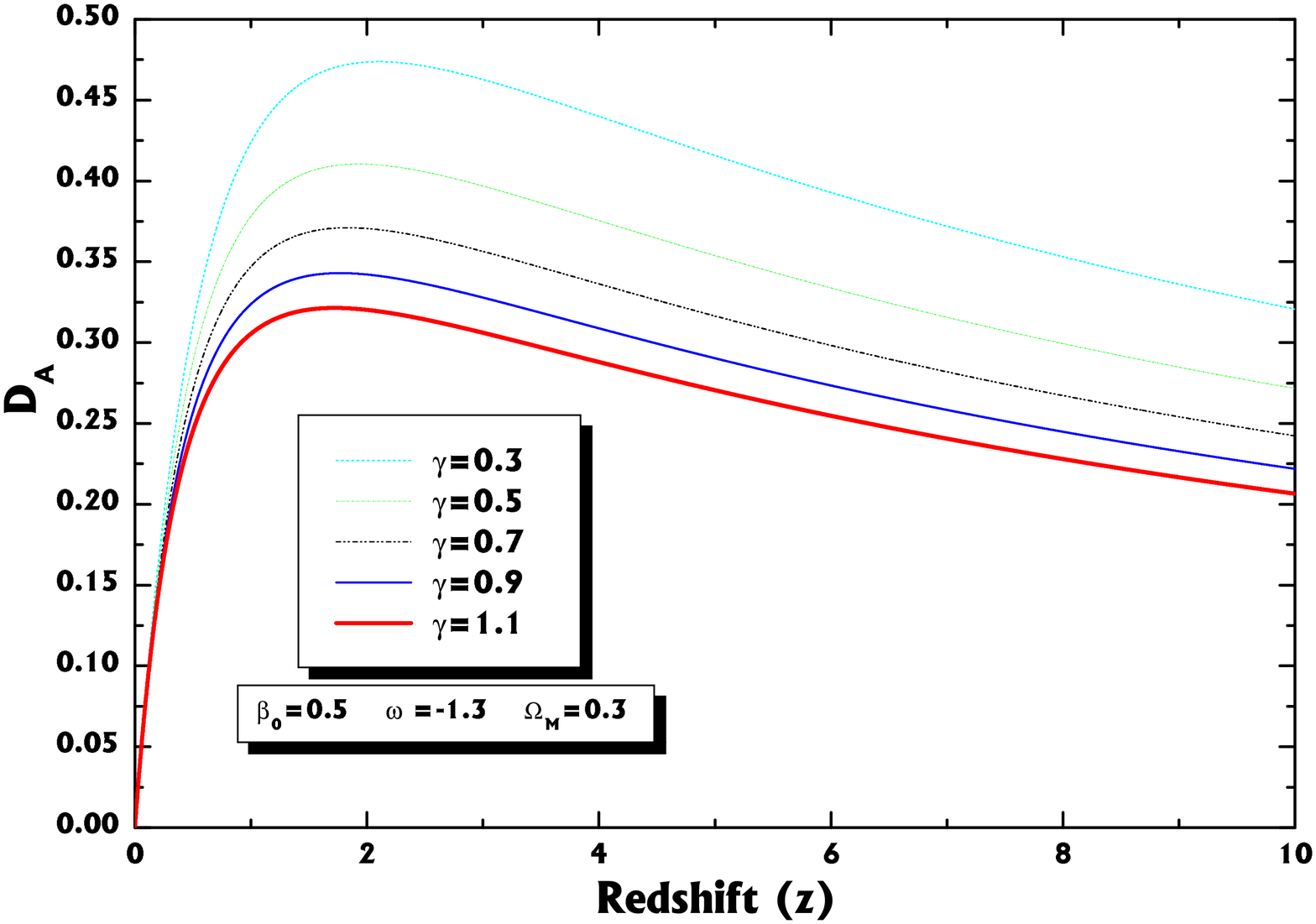,width=4.0truein,height=3.2truein}
\hskip 0.1in} \caption{Effects of the $\gamma$ parameter on the
angular diameter distance. For all curves we fixed $\omega=-1.3$,
$\beta_o = 0.5$ and $\Omega_M=0.3$. Note that the distances increase
for smaller values of $\gamma$.}
\end{figure}

\begin{figure}[t]
\vspace{.2in}
\centerline{\psfig{figure=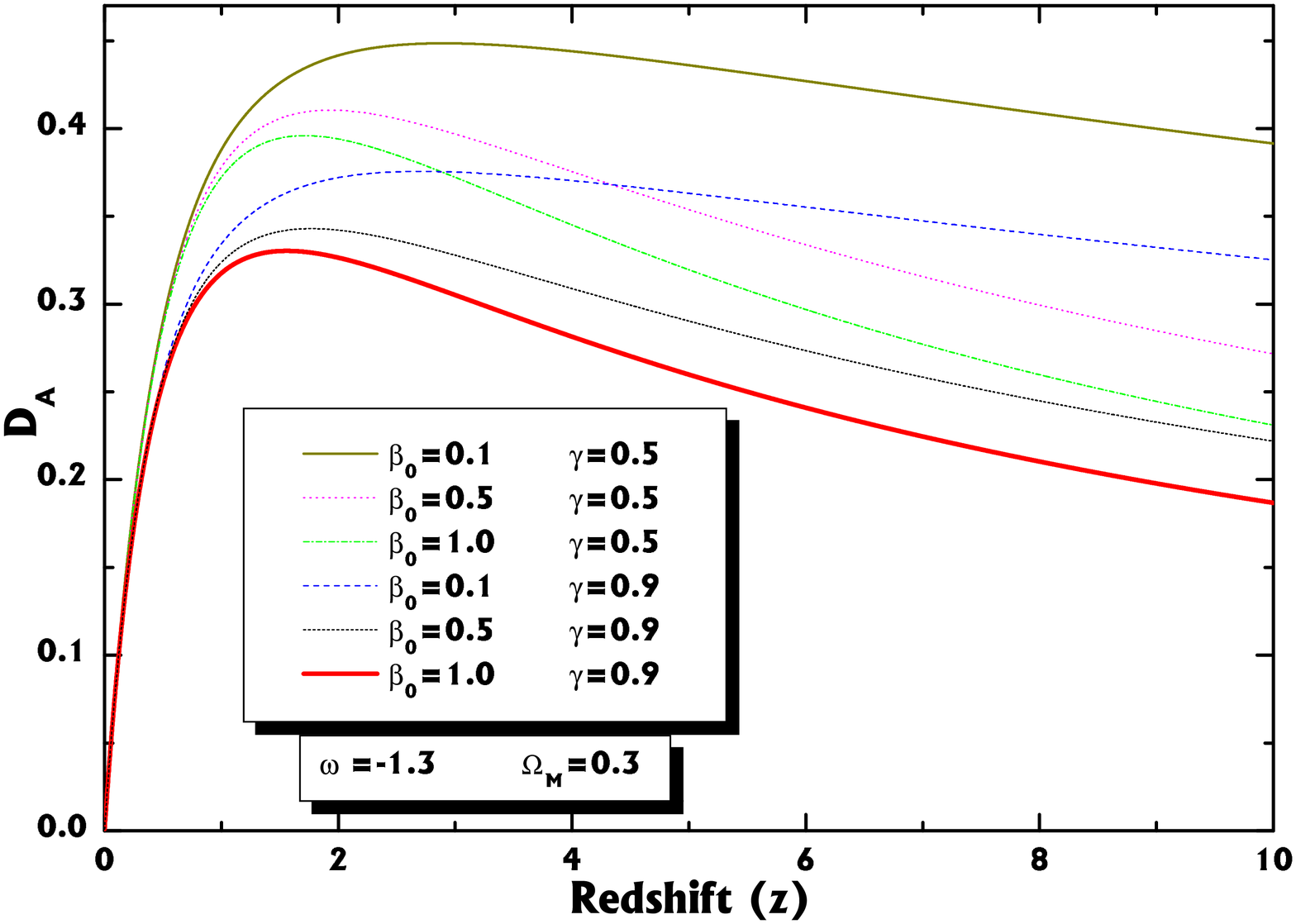,width=4.0truein,height=3.2truein}
\hskip 0.1in} \caption{Influence of the $\beta_o$ parameter on the
angular diameter distance for $\Omega_M=0.3$ and $\omega=-1.3$. The
curves are separated in two sets corresponding to the values of
$\gamma=0.5, 0.9$ as indicated in the box. As expected, both sets
present the same behavior at low redshifts.}
\end{figure}
Here we are concerned with angular diameters of light sources
described as rigid rods and not isophotal diameters. In the FRW
metric, the angular size of a light source of proper length ${\l}$
(assumed free of evolutionary effects) and located at redshift $z$
can be written as
\begin{equation}
\theta(z)= \frac{\ell}{{D_{A}}(z)},
\end{equation}
where $\ell = 100{\l}$h is the angular size scale expressed in
milliarcsecond (mas) while ${\l}$ is measured in parsecs for compact
radio sources (see below).

Let us now discuss the constraints from  angular size measurements
of high $z$ objects on the cosmological parameters. The present
analysis is based on the angular size data for milliarcsecond
compact radio sources compiled by Gurvits {\it et al.} (1999).
This sample is composed by 145 sources at low and high redshifts
($0.011 \leq z \leq 4.72$) distributed into 12 bins with 12-13
sources per bin (for more details see Gurvits et al. 1999). In
Figure 5 we show the binned data of the median angular size
plotted as a function of redshift $z$ to the case with $\gamma=0$
and some selected values of $\Omega_M$ and $\alpha_o = \beta_o/(1
- \beta_o)$ = constant. As can be seen there, for a given value of
$\Omega_M$ the corresponding curve is slightly modified for
different values of the smoothness parameter $\alpha$.

Now, in order to constrain the cosmic parameters, we first fix the
central value of the Hubble parameter obtained by the HST key
project $H_o = 72 \pm 8$ ${\rm{km.s^{-1}.Mpc^{-1}}}$ (Freedman et
al. 2001). Note that this value is greater that the recent
determination by Sandage and collaborators (see astro-ph/0603647),
and it is in accordance with the 3 years release of the WMAP team.
Following standard lines, the confidence regions are constructed
through a $\chi^{2}$ minimization
\begin{equation}
\chi^{2}(l, \omega, \alpha) =
\sum_{i=1}^{12}{\frac{\left[\theta(z_{i}, \l, \omega, \alpha) -
\theta_{oi}\right]^{2}}{\sigma_{i}^{2}}},
\end{equation}
where $\theta(z_{i}$, $\l$, $\omega$, $\alpha)$ is defined from
Eq. (7) and $\theta_{oi}$ are the observed values of the angular
size with errors $\sigma_{i}$ of the $i$th bin in the sample. The
confidence regions  are defined by the conventional two-parameters
$\chi^{2}$ levels. In this analysis, the intrinsic length $\l$, is
considered a kind of ``nuisance" parameter, and, as such, we have
also marginalized over it.
\begin{figure}[t]
\vspace{.2in}
\centerline{\psfig{figure=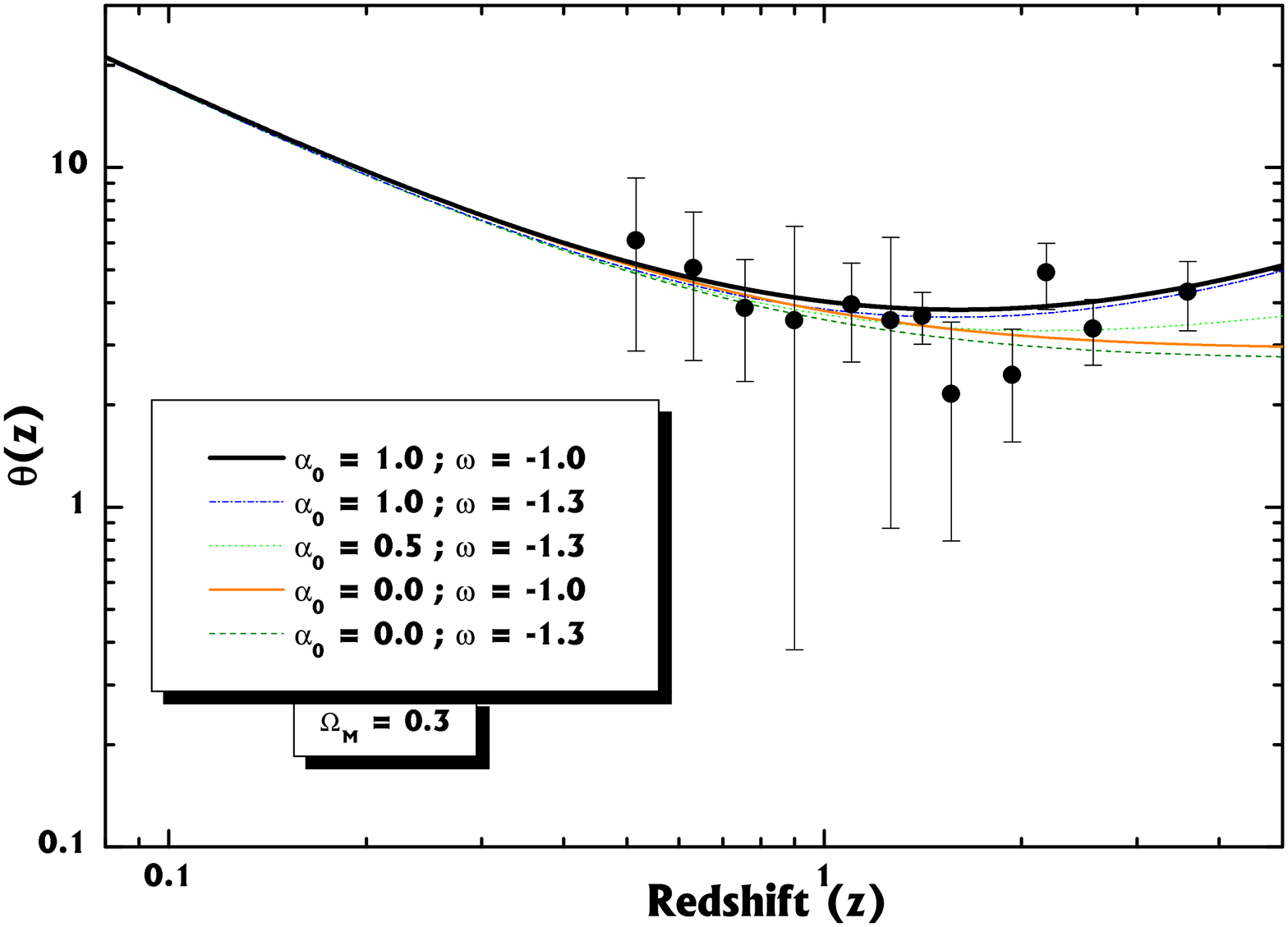,width=4.0truein,height=3.2truein}
\hskip 0.1in} \caption{Angular size versus redshift according to the
ZKDR distance. Curves for $\Omega_M=0.3$, $\gamma=0$ and different
values of $\omega$ are shown. The data points correspond to 145
compact radio sources binned into 12 bins (Gurvits et al. 1999). For
comparison the filled beam $\Lambda$CDM has been included.}
\end{figure}

\begin{figure}[t]
\vspace{.2in}
\centerline{\psfig{figure=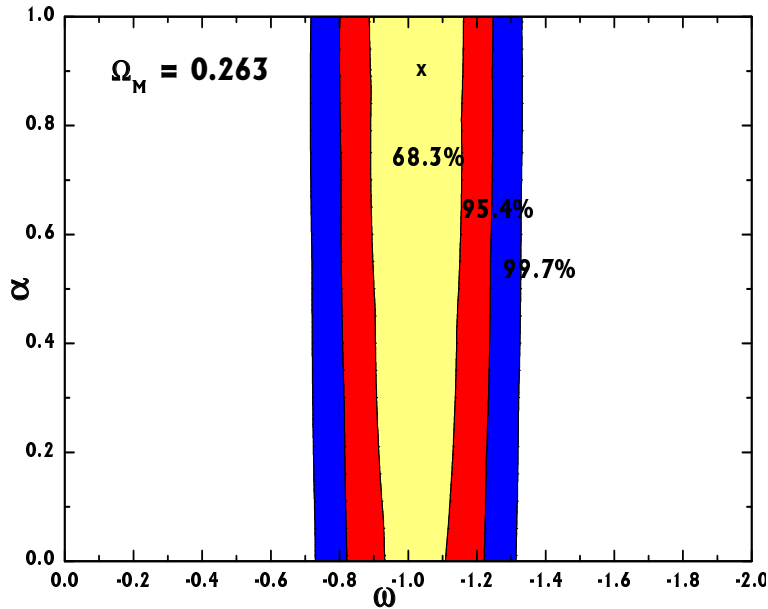,width=4.0truein,height=3.2truein}
\hskip 0.1in} \caption{Confidence regions in the $\omega - \alpha$
plane according to the sample of angular size data by Gurvits et al.
(1999) and fixed $\Omega_M = 0.263$ as shown in panel. The
confidence levels of the contours are indicated. The point ``x"
marks the best fit values, $\omega = -1.03$ and $\alpha = 0.90$.}
\end{figure}

\begin{figure}[t]
\vspace{.2in}
\centerline{\psfig{figure=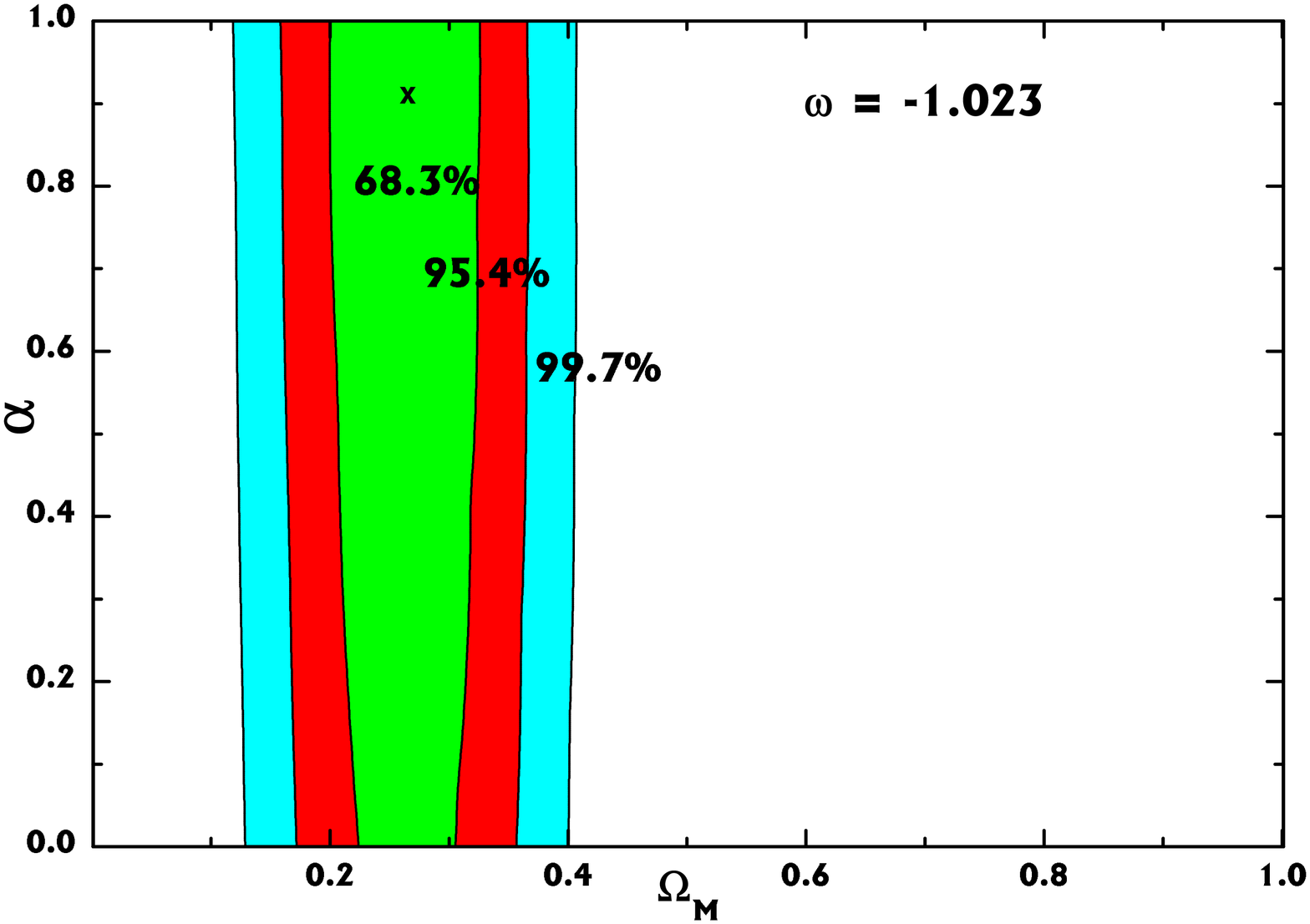,width=4.0truein,height=3.2truein}
\hskip 0.1in} \caption{Confidence regions in the $\Omega_M-\alpha$
plane according to the sample of angular size data by Gurvits et al.
(1999).  For a phantom cosmology with $\omega=-1.2$, the confidence
levels of the contours are indicated. As in Fig. 6, the ``x" also
points to the best fit values shown in the panel.}
\end{figure}
In Fig. 6 we show confidence regions in the $\omega - \alpha$
plane fixing $\Omega_{M} = 0.263$, and assuming a Gaussian prior
on the $\omega$ parameter, i.e., $\omega = -1 \pm 0.3$ (in order
to accelerate the universe). The ``$\times$" indicates the best
fit model that occurs at $\omega = -1.03 $ and $\alpha \simeq
0.9$.

In Fig. 7 the confidence regions are shown in the
$\Omega_M-\alpha$ plane. We have now assumed a Gaussian prior on
$\Omega_{M}$, i.e., $\Omega_{M} = 0.3 \pm 0.1$ from the large
scale structure. From Figs. 6 and 7, it is also perceptible that
while the parameters $\omega$ and $\Omega_M$ are strongly
restricted, the entire interval of $\alpha$ is still allowed. This
shows the impossibility of tightly constraining the smoothness
parameter $\alpha$ with the current angular size data. This result
is in good agreement with the one found by Lima \& Alcaniz (2002)
where the same data set were used to investigate constraints on
quintessence scenarios in homogeneous background, and is also in
line with the one obtained by Barber et al. (2000) who argued in
favor of $\alpha_o = \alpha(z=0)$ near unity (see also Alcaniz,
Lima \& Silva 2004 for constraints on the $\Lambda$CDM model).


\section{Summary and Concluding Remarks}

All cosmological distances must be notably modified whether the
space-time is filled by a smooth dark energy component with
negative pressure plus a clustered dark matter. Here we have
addressed the question of how the angular diameter distance of
extragalactic objects are modified by assuming a slightly
inhomogeneous universe.  The present treatment complements our
previous studies (Alcaniz \& Lima 2000, 2002) by considering that
the inhomogeneities can be described by the
Zeldovich-Kantowski-Dyer-Roeder distance (in this connection see
also, Giovi \& Amendola 2001; Lewis \& Ibata 2002; Sereno {\it et
al.} 2001; Demianski {\it et al.} 2003). The dark energy component
was described by the equation of state $p_x = \omega \rho_x$. A
special emphasis was given to the case of phantom cosmology
($\omega < -1$) when the dominant energy condition is violated.
The effects of the local clustered distribution of dark matter
have been described by the ``smoothness" phenomenological
parameter $\alpha (z)$, and a simple argument for its functional
redshift dependence was given in the Appendix A (see also Figure
1).

The influence of the dark energy component was quantified by
considering the angular diameters for sample of milliarcsecond
radio sources (Fig. 5) as described by Gurvits {\it et al.}
(1999). By marginalizing over the characteristic angular size $l$
and assuming a Gaussian prior on the matter density parameter,
i.e., $\Omega_M = 0.263 \pm 0.07 $(stat) $\pm 0.032$ (sys), the
best fit model occurs at $\omega =-1.03$ and $\alpha = 0.9$. This
phantom model coincides with the central value recently determined
by the Supernova Legacy Survey (Astier {\it et. al.} 2006). On the
other hand, fixing $\omega = -1.3$ and assuming a Gaussian prior
for $\Omega_{M}$, that is, $\Omega_{M} = 0.3 \pm 0.1$, we obtained
the best fit values ($ \Omega_{M} = 0.29$, $\alpha = 0.9$).

Finally, it should be stressed that measurements from the angular
size combined with the ZKDR approach may provide an important and
more rigorous cosmological test in the near future. However, it is
necessary a statistical study for determining the intrinsic length
of the compact radio sources in order to improve the present
results.
\appendix
\section{On the redshift dependence of $\alpha(z)$}

In this Appendix we discuss the functional redshift dependence of
the smoothness parameter, $\alpha(z)$, adopted in this work. By
definition
\begin{equation}
\alpha(z) = 1 - \frac{\rho_{cl}(z)}{\rho_m(z)},
\end{equation}
where $\rho_{cl}$ denotes the clumped fraction of the total matter
density, $\rho_m$, present in the considered FRW type Universe.
This means that the ratio between the homogeneous ($\rho_h$) and
the clumped fraction can be written as $\rho_h/\rho_{cl} =
\alpha(z)/1- \alpha(z)$. How this ratio depends on the redshift?
In this concern, we first remember that $\alpha(z)$ lies on the
interval [0,1]. Secondly, in virtue of the structure formation
process, one expects that the degree of homogeneity must increase
for higher redshifts, or equivalently, the clumped fraction should
be asymptotically vanishing at early times, say, for $z \geq 100$.
This means that $\alpha (z) \rightarrow 1$ at high z. At present,
($z=0$), this fraction may have an intermediate value, say,
$\beta_o$. In addition, it is also natural to suppose that the
redshift dependence of the total matter density, $\rho_m$, must
play an important role in the evolution of their fractions. In
this way, for the sake of generality, we will assume a power law
\begin{equation}
\frac{ \rho_h}{\rho_{cl}} \equiv \frac{\alpha(z)}{1- \alpha(z)} =
\beta_o (\frac{\rho_m}{\rho_o})^{\gamma}.
\end{equation}
where $\beta_o = (\rho_h/\rho_{cl})_{z=0}$ and $\gamma$ are
dimensionless numbers. Finally, inserting $\rho_m (z)$, and solving
for $\alpha(z)$ we obtain:
\begin{equation}
\alpha(z) = \frac{\beta_o(1 + z)^{3\gamma}}{1 + \beta_o(1 +
z)^{3\gamma}},
\end{equation}
which is the expression adopted in this work (see Eq. (9)).

As one may check, for positive values of $\gamma$, the smoothness
function (A.3) has all the physically desirable properties above
discussed. In particular, the limit for high values of $z$ does
not depend on the values of $\beta_o$ and $\gamma$ (both of the
order of unity). Note also that if the clumped and homogeneous
portions are contributing equally at present ($\beta_o=1$), we see
that $\alpha(z=0) = 1/2$ regardless of the value of $\gamma$.
Figure 1 display  the general behavior of $\alpha(z)$ with the
redshift for different choices of $\beta_o$ and $\gamma$. The
above functional dependence should be compared with the other ones
discussed in the literature (Linder 1988, 1998; Campos \& de Souza
2004 and Refs. therein). One of the most interesting features of
(A.3) is that its validity is not restricted to a given redshift
interval.

\acknowledgments
\section*{Acknowledgments}
The authors would like to thank A. Guimar\~aes and J. V. Cunha for
helpful discussions. RCS thanks CNPq No. 15.0293/2007-0 and JASL thanks CNPq and FAPESP
grant No. 04/13668.


\begin{thebibliography}{}

\bibitem{sahni2004} Alam, A., Sahni, V. \& Starobinsky, A. A. 2004, JCAP 0406, 008
\bibitem{alcaniz99} Alcaniz, J. S. \& Lima, J. A. S. 1999, ApJ 521,
L87; {\bf ibdem} 2001, ApJ 550, L133
\bibitem{AL04} Alcaniz, J. S. 2004, PRD 69, 083521
\bibitem{AL05} Alcaniz, J. S. \& Lima, J. A. S. 2005, ApJ 618, 16
\bibitem{AL041} Alcaniz, J. S., Lima,  J. A. S. \&  Silva, R. 2004, IJMPD 13, 1309
\bibitem{Ara07}Ara\'ujo M. E. \& Stoeger W. R. 2007, {\bf arXiv:0705.1846}
\bibitem{Alen04} Allen, S. W. {\it et al.} 2004, MNRAS 353, 457
\bibitem{Asa98} Asada, H. 1998, ApJ 501, 473
\bibitem{Ast05} Astier, P. {\it et al.} 2006, A\&A 447, 31
\bibitem{Bert86} Bertolami, O. 1986, N. Cim. B 93, 36
\bibitem{11} Barber, A. J.  {\it et al.} 2000,  MNRAS 319, 267
\bibitem{bento02} Bento, M. C., Bertolami, O. \& Sen, A. A. 2002, PRD 66, 043507
\bibitem{bilic}Bil\'{\i}c, N., Tupper, G. B., \& Viollier, R. D. 2002, Phys.
Lett. B 535, 17
\bibitem{Bos00} Boisseau, B., Esposito-Farese, G., Polarski, D. \&
Starobinski, A. 2000, Phys. Rev. Lett. 85, 2236
\bibitem{Cal98} Caldwell, R. R., Steinhardt, P. J. 1998, PRD 57, 6057
\bibitem{Cal03} Caldwell, R. R., Kamionkowski, M., Weinberg, N. N. 2003, PRL 91, 071301
\bibitem{CdS04} Campos, M. \& de Souza, J. A. 2004, A\&A 422, 401
\bibitem{carvalho} Carvalho, J. C., Lima, J. A. S. \& Waga, I. 1992,
PRD 46, 2404
\bibitem{chiba97} Chiba, T., Sugiyama, N. \& Nakamura, T. 1997, MNRAS
289, L5
\bibitem{chi00} Chiba, T., Okabe, T., Yamaguchi, M. 2000, PRD 62, 023511
\bibitem{chPa05a} Choudhury, T. R., Padmanabhan, T.  2004, PRD 69, 064033
\bibitem{chPa05} Choudhury, T. R., Padmanabhan, T.  2005, ASP Conference Series
342, 497; 2005, A\&A 429, 807
\bibitem{CoAlMa07}Costa, F. E. M., Alcaniz J. S. \& Maia J. M. F.
2007, arXiv:0708.3800
\bibitem{Co05}Covone, G., Sereno, M. \& de Ritis, R. 2005, MNRAS
357, 773
\bibitem{CS04} Cunha, J. V., Santos, R. C. 2004, IJMPD 13, 1321
\bibitem{Da07}D\c{a}browski, M. P. 2007, {\bf arXiv:gr-qc/0701057}
\bibitem{PH07} de Freitas Pacheco, J. A., Hovarth, J. 2007, {\bf arXiv:0709.1240}
\bibitem{7} Demianski, M., de Ritis, R., Marino, A. A., Piedipalumbo, E. 2003, A\&A 411, 33
\bibitem{1} Dyer, C. C. \&  Roeder, R. C. 1972, ApJ 174, L115;
1972, ApJ 180, L31
\bibitem{Ef02} Efstathiou, G.  {\it et. al.} 2002, MNRAS 330, L29
\bibitem{faraoni02} Faraoni, V. 2002, IJMP D 11, 471
\bibitem{F01} Freedman,  W. {\it et al.} 2001, ApJ  553, 47
\bibitem{FRE87} Freese, K., Adams, F. C., Frieman, J. A., \& Mottola, E.
1987, Nucl. Phys. B 287, 797
\bibitem{6} Giovi, F. \&  Amendola, L. 2001, MNRAS 325, 1097
\bibitem{PG05} Gonz´alez-Diaz, P.F. and Siguenza, C.L., 2004, Nucl. Phys. B697, 363 
\bibitem{G2004} Gurvits, L. I. 2004, New Astron. Rev. 48, 1511
\bibitem{GK99} Gurvits, L. I.,  Kellermann, K. I. \& Frey, S. 1999  A\&A 342, 378
\bibitem{HElis82}Hawking, S. W. \& Ellis, G. F. R. 1973, The large
scale structure of space-time, Cambridge UP, Cambridge
\bibitem{kamen} Kamenshchik, A., Moschell, U., \& Pasquier, V. 2001, Phys.
Lett. B 511, 265
\bibitem{3} Kantowski, R. 1969, ApJ 155, 89
\bibitem{K03}Kantowski, R. 2003, PRD 68, 123516
\bibitem{koc02} Kochanek, C. S. 2002, ApJ 578, 25
\bibitem{koc03} Kochanek, C. S. \&  Schechter, P. L. 2003, {\bf astro-ph/0306040}
\bibitem{LI02} Lewis, G. F.  \& Ibata, R. A. 2002, MNRAS 337, 26
\bibitem{LiViJa03} Lima, J. A. S., Cunha, J. V. \& Alcaniz, J. S. 2003, PRD 68, 023510, {\bf astro-ph/0303388} 
\bibitem{LiMa94} Lima, J. A. S. \& Maia, J. M. F. 1994, PRD 49, 5597
\bibitem{LiTr96} Lima, J. A. S. \& Trodden, M. 1996, PRD 53, 4280
\bibitem{Lima04} Lima, J. A. S. 2004, Braz. Jour. Phys. 34, 194 {\bf astro-ph/0402109}
\bibitem{LA00a} Lima, J. A. S. \& Alcaniz, J. S. 2000, A\&A 357, 393; {\bf ibdem} 2000 Gen. Relativ. Gravit. 32, 1851
\bibitem{LA02} Lima, J. A. S. \& Alcaniz, J. S. 2002, ApJ 566, 15
\bibitem{LA02} Lima, J. A. S. \& Alcaniz, J. S. 2004, Phys. Lett. B 600, 191, {\bf astro-ph/0402265} 
\bibitem{Lin88} Linder, E. V. 1988, A\&A  206, 190
\bibitem{Lin98} Linder, E. V. 1998, ApJ 497, 28
\bibitem{Lin07}Linder, E. V. 2007, {\bf arXiv:0704.2064}
\bibitem{NePe04} Nesseris, S., Perivolaropoulos, L. 2004, PRD 70,  043531
\bibitem{OvCo98} Overduin, F. M. \& Cooperstock, F. I. 1998, PRD 58, 043506
\bibitem{OzTa87} \"{O}zer, M. \& Taha, M. O. 1986, Phys. Lett. B 171, 363; {\bf ibdem}; 1987, Nucl. Phys. B 287, 776
\bibitem{PAD} Padmanabhan, T. 2003, Phys. Rept. 380, 235
\bibitem{PC03} Padmanabhan, T. \&  Choudhury, T. R. 2003, MNRAS, 344, 823
\bibitem{PR03} Peebles, P. J. E. \&  Ratra, B. 2003, Rev. Mod. Phys. 75, 559
\bibitem{perm98} Perlmutter, S. {\it et al.} 1998, Nature 391, 51
\bibitem{peri05} Perivolaropoulos, L. 2005, PRD 71, 063503
\bibitem{peri07} Nesseris, S. \& Perivolaropoulos, L. 2007, JCAP 0701,
018
\bibitem{perrotta21} Perrotta, F. {\it et al.} 2002, MNRAS 329, 445
\bibitem{Piao04} Piao, Yun-Song; Zhang, Yuan-Zhong 2004, PRD 70, 063513
\bibitem{Ries98} Riess, A. G. {\it et al.} 1998, AJ 116, 1009
\bibitem{Ries04} Riess, A. G. {\it et al.} 2004, ApJ 607, 665
\bibitem{Ries07} Riess A. G. {\it et al.}, 2007, ApJ 659, 98
\bibitem{San06} Sandage, A. et al. 2006, {\bf astro-ph/0603647}
\bibitem{janilo05} Santos, J. \& Alcaniz, J. S.
 2005, Phys. Lett.
B 619, 11
\bibitem{SEF} Schneider, P., Ehlers, J. \& Falco, E. E. 1992, {\it Gravitational
lenses}, Springer\,-\,Verlag, Berlin
\bibitem{SPS01} Sereno, M., Covone, G., Piedipalumbo, E., de Ritis, R. 2001, MNRAS
327, 517
\bibitem{SPS02} Sereno, M., Piedipalumbo,  E. \&  Sazhin, M. V. 2002, MNRAS 335, 1061
\bibitem{Shani03} Shani, V. \& Shtanov, Y. 2002, IJMP A 11, 1
\bibitem{Shan03} Shani, V. \& Shtanov, Y., 2003 JCAP 0311, 014,
{\bf astro-ph/0202346}
\bibitem{MiTu07}  Turner M. S., Huterer D., 2007, arXiv:0706.2186
\bibitem{W03} Torres, L. F. B. \& Waga, I 1996, MNRAS, 279, 712
\bibitem{turner97} Turner, M. S., \& White, M. 1997, PRD 56, R4439
\bibitem{wan99} Wang, Y. 1999, ApJ. J. 525, 651
\bibitem{wu07} Wu S.-F., Chatrabhuti A., Yang G.-H., Zhang P-M 2007,
arXiv:0708.1038
\bibitem{2} Zeldovich, Ya. B. 1964, Sov. Astron. 8, 13


\end{thebibliography}
\end{document}